\documentclass[conference]{IEEEtran}
\usepackage{url}

\ifCLASSOPTIONcompsoc
  \usepackage[nocompress]{cite}
\else
  \usepackage{cite}
\fi
\ifCLASSINFOpdf
\else
\fi

\usepackage{graphicx}

\begin{document}
\title{Sustainable Adaptive Security}

\author{
\and
{\rm Liliana Pasquale}\\
University College Dublin
\and
{\rm Kushal Ramkumar}\\
University College Dublin
\and
{\rm Wanling Cai}\\
Trinity College Dublin
\and
{\rm John McCarthy}\\
University College Cork
\and
{\rm Bashar Nuseibeh}\\
University of Limerick
\and
{\rm Gavin Doherty}\\
Trinity College Dublin}

\author{\IEEEauthorblockN{Liliana Pasquale\IEEEauthorrefmark{1},
Kushal Ramkumar\IEEEauthorrefmark{1},
Wanling Cai\IEEEauthorrefmark{2}, 
John McCarthy\IEEEauthorrefmark{3},
Gavin Doherty\IEEEauthorrefmark{2},
Bashar Nuseibeh\IEEEauthorrefmark{4}}
\IEEEauthorblockA{\IEEEauthorrefmark{1} Lero@University College Dublin}
\IEEEauthorblockA{\IEEEauthorrefmark{2} Lero@Trinity College Dublin}
\IEEEauthorblockA{\IEEEauthorrefmark{3} Lero@University College Cork}
\IEEEauthorblockA{\IEEEauthorrefmark{3} Lero@University of Limerick}}

%Telephone: (800) 555--1212, Fax: (888) 555--1212}


\maketitle

\begin{abstract}
With software systems permeating our lives, we are
entitled to expect that such systems are secure by design, and that
such security endures throughout the use of these systems and
their subsequent evolution.
%Limitations of adaptive security systems
Although adaptive security systems have been proposed to continuously protect assets from harm, they can only mitigate threats arising from  changes foreseen at design time.
%
%These approaches focus on threats determined by changes foreseen at design-time, and have not addressed evolving threats brought by unexpected situations, such as unknown vulnerabilities or ineffective security controls.
%
%
In this paper, we propose the notion of \emph{Sustainable Adaptive Security} (SAS) which reflects such enduring protection by augmenting adaptive security systems with the capability of mitigating newly discovered threats.
To achieve this objective, a SAS system should be designed by combining automation (e.g., to discover and mitigate security threats) and human intervention (e.g., to resolve uncertainties during threat discovery and mitigation).
%On the one hand, a SAS  system should be able to  discover and mitigate security threats  autonomously. On the other hand, human intervention can help the system endure security.
In this paper, we use a smart home example to  showcase how we can engineer the activities of the MAPE (Monitor, Analysis, Planning, and Execution) loop of systems satisfying sustainable adaptive security. We suggest that using anomaly detection together with abductive reasoning can help discover new threats and  guide the evolution of security requirements and controls. We also exemplify situations when humans can be involved in the execution of the activities of the MAPE loop and discuss the  requirements to engineer human interventions.

%in a dynamically changing security theatre. %In the context of
%long-lived cyber-physical systems, malicious threats extend over a
%wider attack surface, and the systems must endure for prolonged
%periods of time. 

\end{abstract}


% \fi
\IEEEpeerreviewmaketitle

\section{Introduction}

Security threats are on the rise. Many recent critical cyber
security incidents, such as log4j and SolarWinds, arose from
newly discovered threats~\cite{checkpoint.2022}. % and the number of zero-day
%vulnerabilities reported is also increasing [2], [3]. 
Therefore,
there is a need to build systems that are secure by design, but
that can also detect and mitigate newly discovered security
threats over extended periods of time. Although adaptive security approaches~\cite{Yuan.AutoAS.2014,Tziakouris.ACMSurveys.2018} have
been proposed to mitigate evolving security threats,
they  only address threats arising from changes foreseen at design time. 
However, unanticipated changes, such as newly discovered assets and vulnerabilities or wrong domain assumptions, can bring new security threats and require  security requirements and controls to evolve at runtime.
%Similarly, previous work~\cite{Bergmann.CAISE.2011, Burger.2015} on security requirements evolution evolves security requirements based on pre-defined rules, only when security properties are violated.

In this paper, we propose the notion of \emph{Sustainable
Adaptive Security} (SAS) which reflects the capability of adaptive security systems to preserve security requirements throughout their use and
 subsequent evolution.  Systems satisfying sustainable adaptive security (hereafter referred to as SAS systems) should be capable of discovering
changes that may bring unanticipated security threats and
managing the evolution of security requirements and controls to mitigate such threats. Although autonomy is a desired property of SAS systems, human intervention can be beneficial to preserve security requirements in the long run, for example, by 
monitoring security-relevant data and  supporting decision making~\cite{Gornitz.AISec.2009, Yang.IWC.2018, Fransen.TNO}.

%human intervention canbe helpful, for example, to resolve uncertainties during  to resolve uncertainties durit can also help such systems endure security  in the long run. 

We use an example of a
smart home to motivate the need for sustainable security. We explain how we can engineer the activities of the
MAPE (Monitor, Analysis, Planning, and Execution) loop~\cite{Lara.2019} of SAS
systems. 
We showcase that combining anomaly detection with abductive reasoning can help discover new security threats
and guide the evolution of security requirements and controls to mitigate such threats.
We also exemplify how humans can assist the SAS system in executing the activities of the MAPE loop. Using our example, we discuss three requirements (security, trust, and usability) to  be considered when engineering human interventions.  Finally, we conclude with a discussion of a future research agenda for the research community 
to develop SAS systems.

%The dynamic nature of cyber-physical systems (e.g., new
%devices, changing cyber & physical topologies), evolving
%security goals & requirements, and attacks that impact both
%the cyber and physical space make it more challenging to
%secure them [7]. 

%We then provide that we use to identify the challenges in providing
%sustainable adaptive security. We advocate that sustainable
%adaptive security can be achieved by managing, ensuring and
%preserving evolving security goals and requirements over long
%periods of time. 

%We live in a world where technology influences how we
%interact with the physical space around us. Our smartphones
%operate our smart locks to admit people into our homes, and
%we use smart devices to control our physical environment (e.g.,
%to set a thermostat to the right temperature).

\section{Smart Home Security Example}
\label{sec:example}

\begin{figure}[htpb]
\caption{Smart Home Example}
\label{fig:smartHome}
\centering
\includegraphics[width=0.85\columnwidth]{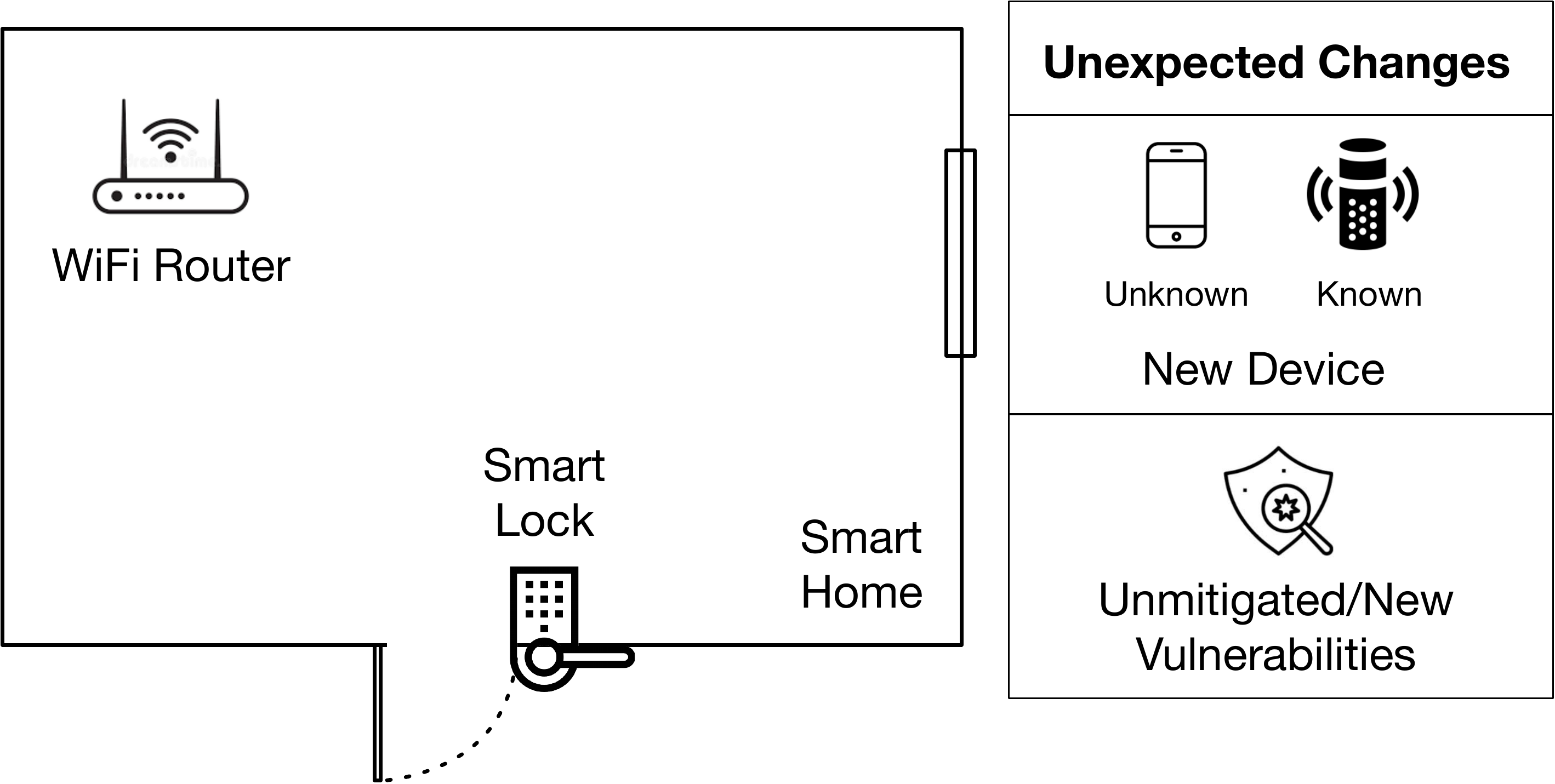}
\end{figure}

 Fig.~\ref{fig:smartHome} illustrates the plan of the smart home (on the left) and exemplifies some of the unexpected changes that can occur (on the right).
The home has a WiFi router and a smart lock which secures the physical space of the home. Devices connected to the WiFi network can potentially control the smart lock~\cite{Veijalainen.2021} and send commands to open and close the door.
A security goal~\cite{Pfleeger.2009} that we aim to maintain is the integrity of the smart home. This goal can be achieved by preventing unauthorised access to the smart home.
Specifically, the smart lock
should only lock or unlock  the door when an authorised user (tenant) intends
for the action to occur. 
%\emph{an outsider should never be inside the house unaccompanied by the house owner}.
 We focus the scope of our research on a system that is currently deployed.
 This is because modern systems include components built by different vendors. Discovering threats during development  may not be feasible because new threats can emerge depending on  how a system is deployed and configured, e.g., depending on the type and location of the  devices in the smart home. 

A \emph{new digital device can connect to the WiFi network}. Such device may be  unknown and can control the smart lock letting an outsider inside  the house unaccompanied. In such case, a new security control should prevent the new device from controlling the smart lock. The presence of unknown devices connected to the WiFi network can indicate security misconfiguration (e.g., weak passwords, insecure encryption protocols) or new vulnerabilities in the WiFi router. Therefore, any security control initially selected by the system will not preserve security in the long run because it may not address the root cause of the problem. %TThe system should also evolve the security requirements and controls to  increase the network protection by, for example, using a stronger password or changing the encryption protocol.  

A new device (e.g., smart speaker) connecting to the WiFi network may, instead, be known and trustworthy and  require temporary authorisation to access the smart lock and other appliances in the smart home. %Because the system cannot observe whether a device is trustworthy, the  tenant may be asked to confirm whether they know the  device. 
However, adding a new device, can also bring new vulnerabilities. For example, a smart speaker may be authorised to control the smart lock, but may be vulnerable to ultrasonic voice command attacks that are inaudible \cite{Mao.IoTJ.2020}.   This not only violates the security requirement from the example, but also puts other devices controllable by the smart speaker at risk. %Thus, the system needs to evolve its security requirements to mitigate vulnerabilities brought by new devices in the smart home. 

If new  devices  connect too frequently to the WiFi network, an adaptive security system can learn that new devices should not be allowed to connect to the WiFi network. The system can also ask the tenant to confirm whether the security control should be enacted.   However, an abrupt  notification without any explanation of why such security control is necessary can  decrease the trust in the system and in the security posture of the smart home. Also, the tenant may not have the expertise to understand whether the suggested security control is effective.

%In this case, the house owner may be required to provide input to indicate whether a device is known and for how long permissions should be granted.

The smart home can also be  vulnerable to attacks exploiting  \emph{unmitigated  or new vulnerabilities}. Some vulnerabilities, although they are known, they are not mitigated due to insufficient security knowledge~\cite{Zeng.SOUPS.2017} or  delayed updates~\cite{Li.CCS.2017}.
%and end-of-life products ~\cite{Cisco.2017}. 
For example, an unmitigated vulnerability in the smart lock (e.g., CVE-2022-32509) may enable man in the middle attacks and allow an outsider inside.  %In this case, information about anomalous traffic or unusual delays when messages are sent/received to/from the smart lock should be used to discover the unmitigated vulnerability and fix it.  %In that case it would be necessary  to ask software/security engineers for advice regarding possible mitigation strategies.
Other vulnerabilities are new and a fix does not exist for them yet.

\section{Sustainable Security}

%So how do we define sustainable security?

In the system security community, sustainable security has been defined as the capability of an organisation to continuously secure a multitude of devices, running potentially outdated versions of software, for long periods of time~\cite{Anderson.CoACM.2018}. It has also been considered as the property of a  network intrusion detection system  to ensure continuous reliable operation~\cite{Koutsandria.CPS-SPC.2015}, or the capability of organisational processes to be more resilient to insider threats~\cite{Hansen.Dagstuhl.2013}. The notion of sustainable security is also related to cyberresilience~\cite{Cyberresilience}, i.e. "the ability to prepare for and recover quickly from both known and unknown threats". Although cyberresilience metrics~\cite{Linkov.2013} have 
been defined to compare system design and prioritise upgrades and maintenance,  previous work has not considered how to engineer systems that can address unknown threats and satisfy evolving security requirements over time, an extension of the concept of sustainability in software
engineering~\cite{Penzenstadler.EASE.2012, Becker.ICSE.2015}.

As systems are increasingly socio-technical, our definition of sustainable security  considers both a system and a human perspective. We define  sustainable security as \emph{the capability of a system to preserve security requirements over sustained periods of time}. To achieve this aim,  systems need to continuously identify and mitigate emerging threats by \emph{adapting} their security controls or evolving the requirements specification. When mitigating threats they should identify short-term  security controls that stop the malicious action, but also long-term controls that address the root-cause of the threat and prevent the recurrence of the same threat. %Since adaptation is an essential characteristic of systems achieving sustainable security, from now on we will refer to such systems as Sustainable Adaptive Security (SAS) systems.

From a human perspective, sustainable security should refer to \emph{the design of technology that enables interactions between the system and the humans to achieve cyberresilience}. To the best of our knowledge, the notion of cyberresilience has not been explored in the HCI community. Previous work has studied how technology plays a role in various forms of resilience to disaster and displacement~\cite{Ahmed.2015,Mark.2008,Mark.2009}, and also how individuals respond to mandated technology adoption~\cite{Karusala.2019}.
Humans should complement the system functionality to support selection and execution of effective security controls that preserve security in the long run. Human participation can be sustained by engineering usable interactions that also increase  trust in the system and its security posture. Usability can be achieved by avoiding obtrusiveness. Trust can be fostered by enabling humans to understand  how the system operates and what is happening at the current moment.

\section{Engineering SAS Systems}

\begin{figure}[htpb]
\caption{Sustainable Adaptive Security System}
\label{fig:SAS}
\centering
\includegraphics[width=0.9\columnwidth]{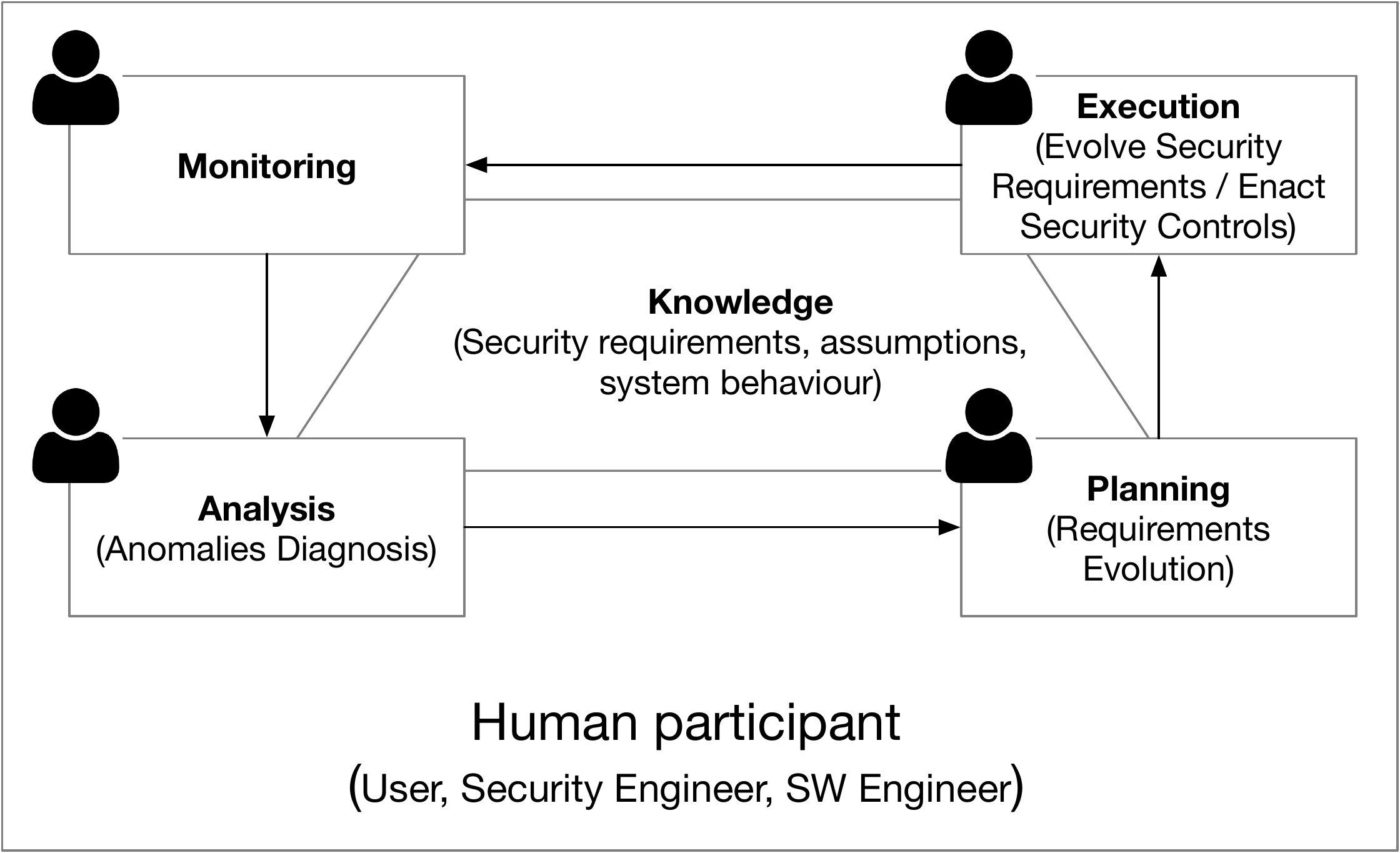}
\end{figure}

We revisit the activities of the MAPE loop to engineer Sustainable Adaptive Security (SAS) systems, as shown in Fig.~\ref{fig:SAS}. 
To discover new threats the monitoring activity not only should assess satisfaction of requirements and assumptions on the operating environment (domain assumptions)  but should also detect unusual behaviours (anomalies), which can provide indicators that security requirements and controls should evolve. We assume that the SAS system can collect data about the user behaviour (e.g., entering, exiting, whether the door is locked or unlocked or whether new devices are present in the network). 
 The analysis activity should  diagnose  unusual behaviour and assess whether new threats can materialise. The planning activity should identify how  security requirements and controls should evolve, if necessary. The execution activity should enact security controls. We assume that  the SAS system is separated from the system under protection (smart home), although it can observe/control the devices in the smart home.
 
Humans can be involved in the execution of the activities of the MAPE loop in capacities other than the traditional user~\cite{Baumer.2015,Baumer.2017} 
such as security and software engineers. For example, during monitoring a user can provide information about data that cannot be observed  by the SAS system (e.g., whether a device connected to the WiFi network is trustworthy). During analysis, information about a discovered anomaly can be provided to software/security engineers to diagnose the anomaly, if it cannot be done automatically. During planning, a user can confirm whether a security control can be enacted (e.g., forcing closure of the door of the smart home). Finally, humans can support execution of security controls (e.g., the tenant may be asked to return home).

\begin{figure*}[htpb]
\caption{Goal Model}
\label{fig:Goals}
\centering
\includegraphics[width=1.8\columnwidth]{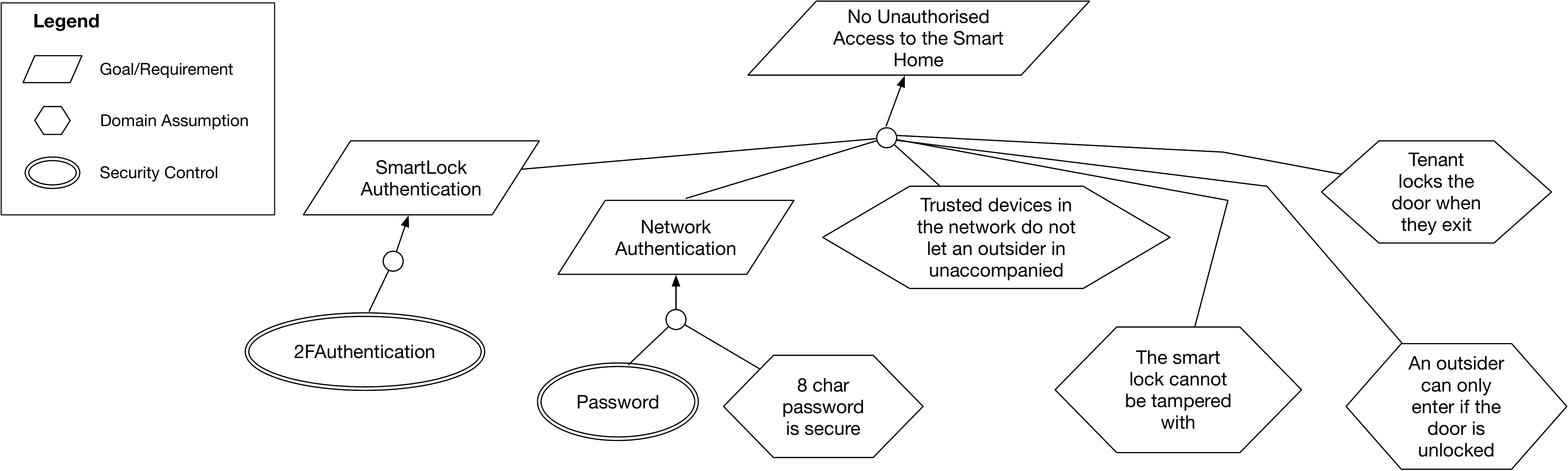}
\end{figure*}

Similarly to previous work on adaptive security~\cite{Salehie.2012}, we use a goal
modeling framework~\cite{VanLamsweerde.2009} to  make security requirements refinements and dependencies on environment conditions explicit. We suggest that a representation of security requirements %and more generally security concerns (assets, threats, vulnerabilities, attack)
can help engineer the activities of the MAPE loop of a sustainable adaptive security system.
%We use our example to show how a goal model can  guide  autonomy and human participation in sustainable adaptive security systems.
Fig.~\ref{fig:Goals} shows a simplified goal model of the smart home example. The root goal represents the security requirement related to the authorised access to the smart home. This requirement can be achieved by authenticating access to the smart lock and securing access to the WiFi network. Two-Factor authentication is used to access the smart lock. An 8 character password is used for the WiFi network and the length of the password is assumed to be sufficient to protect the network. 
We assume that a) trusted devices in the WiFi network do not let an outsider in the smart home unaccompanied; b) the smart lock cannot be tampered with; c) an outsider can only enter if the door is unlocked and d) the tenant always locks the door when they exit.
%This goal can be achieved by ensuring the integrity of the smart lock, under the assumption that the tenant locks the door when they exit and that an outsider cannot enter if the door is locked. The integrity of the smart lock can be achieved by ensuring its physical integrity, and supporting authentication to connect to the smart lock and the WiFi network. 

In the rest of this section we showcase how abductive reasoning can be used to diagnose whether an anomaly can lead to the violation of the authorisation requirement for the unexpected changes described in Section~\ref{sec:example}.  We encoded the goal model and the system functionalities (e.g., opening/closing door, entering/exiting) using Answer Set Programming and use clingo to perform abductive reasoning\footnote{For reasons of space, we omit information about the ASP models in this paper and refer the interested reader to \url{https://tinyurl.com/256tfktc}}. As suggested by Alrajeh et al.~\cite{Dalal.ICSE.2020}, symbolic learning techniques can use traces of the system behaviour satisfying and violating the authorisation requirement to learn how security requirements and controls should evolve.  We also exemplify  how humans can participate in the activities of the MAPE loop.

\subsection{New Device Connects to the WiFi Network}
The monitoring activity detects that a new device is connected to the WiFi network. However, to diagnose the anomaly the SAS system needs to know whether the device is trustworthy, i.e. known to the tenant. 
If the device (\texttt{d1}) is not trustworthy, the ASP model will be updated accordingly and will generate a trace showcasing how the authorisation requirement  can be violated, as shown  below.

\begin{small}
\begin{verbatim}
exit(tenant,home,1), close(sl,2), 
   open(d1,sl,3), enter(outsider,home,4), 
      in(outsider,home,4).
\end{verbatim}
\end{small}

\noindent After the tenant exits at time 1 and closes the door at time 2, the new device sends a command to the smart lock (\texttt{sl}) to open the door and lets and outsider inside. 

The SAS system can learn to evolve the specification by, for example,  preventing the new device  from sending an open command to the smart lock as shown below.

\begin{small}
\begin{verbatim}
X!=d1 :- open(X,sl,T), net_device(X), T = 0..4.
\end{verbatim}
\end{small}

In this case, human participation can be sought to understand whether the security control selected by the SAS system should be enacted. To foster trust, observability and transparency principles should be satisfied~\cite{Gil.2019}. In other words, the SAS system should indicate that a network device that is not trustworthy was detected (observability) and that the suggested security control is necessary to prevent the new device from letting an offender inside unaccompanied (transparency). Alternatively, it may be desirable to evolve security controls differently, for example by removing the new device from the WiFi network. Although alternative security controls can be discovered automatically by the system, the tenant may not have the expertise to select one appropriately. Thus,  intervention of the system/security engineer may be necessary to perform this task. If an engineer is required to modify security controls, the SAS system should also satisfy the intelligibility principle~\cite{Gil.2019} by communicating in what ways security controls can be modified without introducing additional vulnerabilities.

However, the security controls identified above  may not be sustainable if the new device is trustworthy. For example, the tenant may want to use a new smart speaker to send commands to the smart lock to open and close the door. Thus, the SAS system should ask the tenant whether the device is trustworthy, to avoid selecting ineffective security controls. To increase the likelihood that the tenant completes the task successfully, the SAS system should satisfy the feedforward principle~\cite{Gil.2019} by providing the tenant with information about the detected device (e.g., type of device). Similarly to the previous example,  to satisfy observability, the SAS system should indicate that a new device was detected. To satisfy transparency the SAS system should indicate that,  if the device is trustworthy, access to the network and the home appliances will be granted.

Now, let us imagine that new devices keep connecting to the WiFi network frequently. 
 To avoid sending frequent notifications to the tenant, the SAS system may learn that new devices should not be allowed in the WiFi network by default or  can ask the tenant whether, from now on, they may want to prevent any new device from connecting to the network. In this case, the SAS system should  satisfy observability by indicating that new devices are connected to the network too frequently. It should also satisfy transparency by explaining  that forbidding access to new devices will require the tenant to manually grant access to the network when a new device will connect to the WiFi network in the future.

 However, these security controls may not be sustainable, because they may not address the root cause of the problem, which can be that the network authentication is not effective. 
 Thus, the SAS can use the structure of the goal model to identify which security control is responsible to regulate access to the network and, for example, evolve the domain assumption related to the password strength by increasing the minimum number of characters for a password to be considered strong. Changing the domain assumption will force the system to learn that a longer password needs to be enforced. In this case, human input should be sought to identify the minimum number of characters that a password should have.  Since the tenant  may not have the expertise to advise on the password length, a security/software engineer  may need to be involved to perform this task.

\subsection{New Vulnerabilities}
If the monitoring activity measures the latency of requests and responses to and from  appliances in the smart home it can discover unusual latencies that can indicate the presence of a man-in-the-middle attack. These types of attacks can be due to authentication vulnerabilities. For example, the Nuki Smart Lock was found to lack SSL/TLS certificate validation, allowing an attacker to perform a man-in-the-middle attack and intercept network traffic (CVE-2022-32509).

Such vulnerability cannot be diagnosed automatically and the SAS system needs to identify a security control that could prevent an attack from happening. In such a case, the domain assumption on the trusted devices in the network may no longer be valid. Removing that domain assumption from the ASP model makes the system identify a violating trace where a trusted device in the network could still let an outsider inside unaccompanied. This trace can be used to learn a security control that blocks any incoming traffic to the smart lock. This security control is not sustainable in the long run because it does not address the root cause of the problem. 
Intervention of a security/software engineer needs to be sought to confirm the presence of a vulnerability. If information about the vulnerability has already been disclosed (e.g., a CVE entry is available) a patch needs to be identified and applied to the vulnerable device. Also, the requirement specification needs to be updated with an indication of the vulnerability and the fix. Alternatively, advice needs to be sought from the software/security engineer on how the disclosure of a new vulnerability and how to fix it.

\section{Related Work}
\label{sec:RelWork}

%Requirements Evolution
Previous work on security requirements engineering~\cite{Liu.2003, vanLamsweerde.2004, Giorgini.2005,Haley.SE.2008, Yu.2015,Pasquale.2016,Sven.RE.2017} has focused on modelling and reasoning about security concerns to assess risks, identify security controls and estimate the impact of design choices on the protection of the critical assets
of the system. These approaches typically require a complete model of the system and have not been designed to support requirements evolution.
Other approaches~\cite{Bergmann.CAISE.2011,Burger.2015} trigger requirements evolution according to pre-defined rules, when security properties are violated. 
More recently, Drodzov et al.~\cite{Drozdov.TPS} have used symbolic-based learning to learn security policies from historical data produced by anomaly detectors. The authors adopt a domain-specific function to guide the learning process towards the best policies for anomaly detection. Although promising, this work has not been adopted to evolve security requirements at runtime.

%Adaptive Systems
%Adaptive security approaches attempt to mitigate threats by dynamically applying security controls to a system to continuously satisfy its security goals~\cite{Salehie.2012}. However, surveys on adaptive security have shown that previous work in this space is not suited to addressing newly discovered threats~\cite{Yuan.AutoAS.2014, Tziakouris.ACMSurveys.2018,Gheibi.AutoAS.2021}.
In the adaptive security community, previous work has suggested to evolve the representation of threats based on the automated derivation of changing architectural system models from runtime and operational system artifacts~\cite{Landuyt.SEAMS.2021}. %However, this work does not provide a concrete solution for the automated discovery of changes in the architectural system models and the identification of new threats at runtime. 
Calo et al.~\cite{Calo.2018} propose an approach to generate access control policies dynamically when the environment changes, assuming that a set of constraints associated with the resources to be accessed are satisfied. 
To the best of our knowledge, there is no systematic technique to detect and prevent/mitigate evolving threats brought by unexpected  changes, such as vulnerabilities and ineffective security controls at runtime. 
%Previous work on network security has extensively explored techniques to detect newly discovered threats~\cite{Cao.INFOCOM.2017, Casas.CC.2012, Serror.ARES.2018}. 
%However, these techniques do not attempt to mitigate the threats identified, or incorporate human intervention to perform decision making or enact security controls. %Previous work aimed at detecting new cyber-physical threats~\cite{Li.II.2021, FreitasdeAraujo-Filho.IoTJ.2021} also does not apply security controls or seek human intervention, and surveys on cyber-physical systems security find the need for further research in this regard~\cite{Mitchell.ACMSurveys.2014, Luo.ACMSurveys.2021}.

Explanations have been used  to involve humans in the execution of the activities of an adaptive system~\cite{Welsh.2014,Li.2020}. In the security domain, Nhlabatsi et al.~\cite{Nhlabatsi.2015} help users understand security decisions  by establishing traceability relationships between requirements and security concerns (e.g., threats, attacks and vulnerabilities). More recently, Adepu et al.~\cite{Adepu.TAAS.2022}   characterise explanations in terms of  content, effect, and cost (i.e. human effort necessary to understand the explanation). The authors use probabilistic reasoning to determine when an explanation should be used to improve overall system utility. However, they do not configure the content of an explanation depending on the input required from the human participant. 
Models of human participants  have been used to decide whether humans should be involved during the planning~\cite{Lloyd.2017} or execution~\cite{Camara.2015} activity of adaptation.  Li et al.~\cite{Li.2021} propose a  framework to reason about the usage of preparatory notifications in adaptive systems. Preparatory notifications can focus the attention of human participants  on the task to be performed and help them understand its context before task execution. However, previous work has provided limited guidance on how human intervention should be requested and provided. Gil et al.~\cite{Gil.2019} proposes a conceptual framework to characterise the cooperation between humans and autonomous cyber-physical systems and provide techniques for applying the framework  to design  human integration. However, this framework has not been applied in the context of an adaptive security system.

\section{Research Agenda}
%We identify research challenges aimed to engineer the activities of the MAPE loop of SAS systems that can be performed automatically and with human interventions.
%We now discuss the challenges to be addressed by SAS systems to perform the activities of the MAPE loop automatically and to engineer human interventions.

\subsection{Engineering Autonomy in SAS Systems}

During monitoring SAS systems need to identify  events that can be observed by the devices that can be controlled in the environment and  those for which human intervention is necessary. For example, a combination of device finger-printing and human intervention~\cite{OConnor.WMN.2019} can be used to identify new devices connected to the WiFi network.

During analysis, to detect the root cause of an anomaly, it will be necessary to identify parts of the requirements model that require revision. The devices and domain assumptions that are affected by the anomaly can be used to identify the parts of the model that require revision automatically. In our example, frequent connectivity of new devices to the WiFi network required us to revise the security controls protecting the network. If not possible, human intervention should be sought to select the parts  of the model that need revision. There can also be situations when anomalies cannot be diagnosed, for example, when we have unanticipated or new vulnerabilities.

We will attempt to detect unanticipated vulnerabilities  by referencing the behavioural pattern of the anomaly against an up-to-date knowledge database of attacks created from vulnerability disclosures and vulnerability databases.
To detect and mitigate new  vulnerabilities, it will be necessary to identify what type of information   can be disclosed to software and security engineers to obtain meaningful input.

During planning there can be several ways in which security controls can evolve. This will depend on the examples that are used to perform the learning activity. One challenge to be addressed is the generation of examples and the identification of heuristics that can drive the learning towards the selection of security controls that are effective in the long run, while avoiding human intervention.

In this paper we assumed that the  requirements, functionalities, and domain assumptions of the system to be protected are specified in advance. However, this may not always be possible, for example,  when new components are added to the system at runtime. Thus, it will be interesting to understand how the system functionalities, domain assumptions and security controls can be learnt from scratch. For example, transfer learning can be used to learn the behaviour of new components from similar ones. Online learning can be used to learn how heterogeneous components can interact when placed together. Then, neural symbolic learning~\cite{garcez2022neural} can be used to learn a model of the system behaviour. Finally symbolic learning can be used to learn the security controls that satisfy the system security requirements. Human input will be necessary to identify the security requirements and the security controls available, and  to revise the model learnt from scratch.
 
 %Incident reports can be instantiated on cyber-physical systems to determine if previous incidents can recur~\cite{Alrimwai.SE.2022}.

%\textbf{Ensuring sustainable security.} We intend to investigate the usage of machine learning techniques such as active learning~\cite{Gornitz.AISec.2009, Yang.IWC.2018} that include human inputs in decision making to \textit{detect (un)known security threats}. Since we distinguish between unanticipated and undisclosed threats, the detection mechanisms vary between them. A preliminary step would be to use efficient techniques to create a baseline of the normal behaviour of the devices in a smart space~\cite{Hao.TASE.2021} and then identify anomalous device behaviours in a \textit{timely} manner.  We will explore techniques to identify \textit{cyber-physical threats} that monitor the physical media for attacks~\cite{Mao.IoTJ.2020} in addition to the observed device behaviour. For \textit{threat mitigation}, we intend to explore automated techniques such as network segmentation from the intrusion detection literature, and techniques that require user input such as device security configuration (e.g., enabling voice recognition on a smart speaker, choosing HTTPS instead of HTTP)~\cite{Hammi.CS.2022}. 

\subsection{Engineering Human Interventions}
Engineering human interventions requires identifying the tasks to be performed by human participants. In this paper, we initially identified some of them (e.g., provisioning of monitoring information, hypothesis on the root cause of an anomaly, and selection/modification of security controls). However, it will be necessary to elicit such tasks in a more systematic way using a larger set of scenarios from different application domains (e.g., cloud computing).
Once tasks are determined, it will be desirable to identify the roles and expertise of humans that can perform these tasks. 

Moreover, SAS system should enable effective and usable interaction with humans. To achieve this aim, it will be necessary to consider the following aspects: (1) "who" -- who will be involved in the task, (2) "when" --  human or the system initiate the actions, and the level of automation or human control for performing the task~\cite{Muller.2022} (e.g., some tasks can require selection but others can require modification of security controls); (3) "how" --  human and the system can communicate with each others   (4) what information should be exchanged between the human and the system. To foster trust  the information exchanged should satisfy some of the properties of human-system integration~\cite{Gil.2019}.

%identifying the information that should be exchanged between the human and the SAS system.  %For example, it should describe the current state of the system to satisfy observability, should explain the impact that the action selected by the user will have on the system to satisfy transparency.

Evaluating sustainable adaptive security systems also brings new research challenges. From a system perspective, properties such as timeliness and longevity are very relevant. Timeliness refers to how fast such systems can prevent threats from happening compared to traditional adaptive security systems. Longevity can refer to  how long security requirements are satisfied. To evaluate SAS systems from a human perspective, we can consider assessing the successful completion of the task by the participants, but also evaluate sustainable security from an "experiential perspective" by assessing  the perceived security level of the system experienced by human participants~\cite{Haney.CHI.2020}.

\bibliographystyle{IEEEtran}
\bibliography{biblio}

\end{document}